\documentclass[pre,amssymb,amsmath,preprintnumbers,nofootinbib,showpacs,12pt]{revtex4-1}

\usepackage{graphicx}
\usepackage{subfigure}
\usepackage{amsmath}
\usepackage{color}
\usepackage{verbatim}
\usepackage{epstopdf}
\usepackage[CaptionAfterwards]{fltpage}

\begin{document}

\title{The free energy cost of accurate biochemical oscillations}
\author{Yuansheng Cao$^1$}
\author{Hongli Wang$^1$}
\author{Qi Ouyang$^{1,2}$}
\email{qi@pku.edu.cn}
\author{Yuhai Tu$^{3,2}$}
\email{yuhai@us.ibm.com}

\affiliation{$^1$The State Key Laboratory for Artificial Microstructures
and Mesoscopic Physics, School of Physics, Peking University, Beijing, 100871, China}
\affiliation{$^2$Center for Quantitative Biology
and Peking-Tsinghua Center for Life Sciences, AAIC, Peking University, Beijing, 100871, China}
\affiliation{$^3$IBM T. J. Watson Research Center, Yorktown Heights, New York 10598, USA}



\begin{abstract}
Oscillation is an important cellular process that regulates timing of different vital life cycles. However, in the noisy cellular environment, oscillations can be highly inaccurate due to phase fluctuations. It remains poorly understood how biochemical circuits suppress phase fluctuations and what is the incurred thermodynamic cost. Here, we study four different types of biochemical oscillations representing three basic oscillation motifs shared by all known oscillatory systems. We find that the
phase diffusion constant follows the same inverse dependence on the free energy dissipation per period for all systems studied. This relationship between the phase diffusion and energy dissipation is shown analytically in a model of noisy oscillation. Microscopically, we find that the oscillation is driven by multiple irreversible cycles that hydrolyze the fuel molecules such as ATP; the number of phase coherent periods is proportional to the free energy consumed per period. Experimental evidence in support of this universal relationship and testable predictions are also presented.
\end{abstract}


\maketitle

\clearpage
\section{introduction}
Living systems are dissipative, consuming energy to perform key functions for their survival and growth. While it is clear that free energy \cite{EisenbergHill1985,Hill2005,QianBeard2008} is needed for physical functions, such as cell motility \cite{Julicher1997} and macromolecule synthesis \cite{Nelson2008}, it remains poorly understood whether and how regulatory functions are enhanced by free energy consumption.  The relationship between biological regulatory functions and nonequilibrium thermodynamics has been an active area in biophysics \cite{BialekSetayeshgar2005,HuChen2010,LanSartori2012,LanTu2013,SkogeNaqvi2013,LangFisher2014}. 
For example, recent studies in different cellular adaptation processes demonstrated that the cost-performance trade-off follows a universal relationship, independent of the detailed biochemical circuits \cite{LanSartori2012,LanTu2013}.


Oscillatory behaviors exist in many biological systems, e.g., glycolysis \cite{Goldbeter1996}, cyclic AMP signaling \cite{MartielGoldbeter1987}, cell cycle \cite{Pomerening2003,Tsai2008,FerrellTsai2011}, circadian rhythms \cite{Goldbeter1996,HogeneschUeda2011}, and synthetic oscillators \cite{ElowitzLeibler2000,Stricker2008}. These biochemical oscillations are crucial in controlling the timing of life processes. Much is known now about the structure of biochemical circuits responsible for these oscillatory behaviors. There are a few basic network motifs, illustrated in Figure 1a, which are responsible for all known biochemical and genetic oscillations \cite{Goldbeter1996,MartielGoldbeter1987, FerrellTsai2011,HogeneschUeda2011, ElowitzLeibler2000}. These network motifs share a few essential features, such as nonlinearity, negative feedback, and a time delay, as summarized by Novak and Tyson in \cite{NovakTyson2008}. However,
in small systems such as a single cell, the dynamics of oscillations are subject to large fluctuations from the environment, due to their small sizes. Thus, one may ask how biological systems maintain coherence of oscillations amidst these fluctuations \cite{BarkaiLeibler2000}. Here, we study the thermodynamic cost of controlling oscillation coherence in different representative oscillatory systems and investigate whether there is a general (universal) relation between the accuracy of the oscillation and its minimum free energy cost that may apply to all biochemical oscillations.

We study four specific models, the activator-inhibitor (AI) model, the repressilator model, the brusselator model, and the allosteric glycolysis model, chosen to exemplify the three different basic oscillation motifs, as shown in Fig. 1. For all the systems studied, a finite (critical) amount of free energy is needed to drive them to oscillate. Beyond the onset of oscillation, extra free energy dissipation is used to reduce the phase diffusion constant and thus enhance the coherence time and phase accuracy of the oscillations. 
A general inverse relationship between the phase diffusion constant and the free energy dissipation is found in all the four models studied, suggesting that the relation may hold true for all biochemical oscillations. The energy-accuracy relation for noisy oscillations is also verified analytically in the noisy complex Stuart-Landau equation. In the following, we report these results followed by a in-depth discussion of a plausible general microscopic mechanism/strategy for energy-assisted noise suppression.


\section{models and results}

\subsection{Four biochemical oscillators representing the basic network motifs}

\noindent All known biochemical and genetic oscillators contain at least one of the basic motifs (or their variance) in network topology \cite{NovakTyson2008,vanDorp2013}. To search for general principles in these noisy oscillatory systems, we study four biochemical systems (Fig. \ref{fig1}), each representing one of the three basic network motifs responsible for oscillatory behaviors. The first one is the activator-inhibitor (AI) system, where a negative feedback is interlinked with a positive feedback (Left panel, Fig. 1a). This regulatory motif is common in biological oscillators, like the circadian clock in cyanobacteria \cite{Nakajima2005,RustMarkson2007},  cell cycle in frog egg \cite{Goldbeter1991,Pomerening2005}, cAMP signaling in {\it Dictyostelium}, and genetic oscillators in synthetic biology \cite{Stricker2008,DaninoMondragon-Palomino2010,Prindle2012}. We implement this motif in a simplified biological network with a phosphorylation-dephosphorylation cycle (Fig. 1b). The second model is a repressilator, which consists of three components connected in a negative feedback loop, such that each component represses the next one in the loop, and is itself repressed by the previous one (Middle panel, Fig. 1a). The first synthetic genetic oscillator was built with this motif \cite{ElowitzLeibler2000}. Many important transcriptional-translational oscillators also use this motif as their backbone, such as circadian clock in mammalian cells \cite{HogeneschUeda2011}, NF-$\kappa$B signaling \cite{KrishnaJensen2006}, and the p53-mdm2 oscillations in cancer cells \cite{Geva-ZatorskyRosenfeld2006}. Here, we take the repressilator composed of CDK1, Plk1, and APC in a cell cycle as our case study (Fig. 1c). The third model we chose is the brusselator, which is one of the simplest two-component systems that can generate sustained oscillations (Right panel, Fig. 1a).  The concentration fluctuations due to small molecule numbers were analyzed in \cite{QianSaffarian2002}, here, we aim to study the effect of noise on the phase of the oscillation. The brusselator (Fig. 1d) is a special kind of substrate-depletion model \cite{SzallasiStelling2006}, where substrate $S$ is converted by a process that is amplified autocatalytically by the product $P$. Examples of substrate-depletion motif are oscillations in glycolysis \cite{GoldbeterLefever1972,Goldbeter1996} and Calcium signaling \cite{Dupont1991}. Here, we examine the noise effect in glycolysis oscillations, where the allosteric enzyme PFK catalyzes substrate to product in a network shown in Fig. 1e.

In our study, we introduced a parameter $\gamma$ to characterize the reversibility of the biochemical networks. In a reaction loop, $\gamma$ corresponds to the ratio of the product of the reaction rates in one direction (e.g., counter-clock-wise) and that in the other direction (e.g., clock-wise). When $\gamma=1$, the system is in equilibrium without any free energy dissipation. For $\gamma\ne 1$, free energy is dissipated. Here, we study the relationship between the dynamics and the energetics of the biochemical networks by varying $\gamma$. The mathematical details of the four models are described in the Supplemental Information (SI), all parameters (e.g., reaction rates, concentrations, time, and volumes) are shown here as dimensionless numbers with their units explained in SI.

\subsection{Phase diffusion reduces the coherence time}

In all four models that we studied, there is an onset of oscillation as $\gamma$ decreases below a critical value $\gamma_c\;(<1)$. This means that a finite critical free energy dissipation ($W_c>0$) is needed to generate an oscillatory behavior (see Fig. S1 in SI). In Fig. 2a, two trajectories of the concentration of the inhibitor $X$ are shown for $\gamma<\gamma_c$ in the activator-inhibitor model, where $\gamma_c=2\times 10^{-3}$. As evident in Fig. 2a, biochemical oscillations are noisy. To characterize the coherence of the oscillation in time, we computed the auto-correlation function $C(t)$ for a given concentration variable $x$ in the network. As shown in Fig. 2b, $C(t)$ follows a damped oscillation:
\begin{equation}
C(t)\equiv \frac{\langle(x(t+s)-\langle x \rangle)(x(s)-\langle x\rangle)\rangle _s}{\langle x^2\rangle-\langle x\rangle ^2}=\exp(-t/\tau_c)\times \cos(2\pi t/T),
\end{equation}
where $T$ is the period and $\tau_c$ defines a coherence time for the oscillation.

The oscillatory state breaks time translation invariance (symmetry) of the underlying biochemical system. As a result, the phase of the oscillation is a soft mode and follows diffusive dynamics in the presence of noise. To quantify the phase diffusion, we simulated many trajectories in the model(s) with the same parameters and the same initial conditions. In Fig. 2c, the peak times for $500$ trajectories in the AI model are shown in a raster plot together with the peak time distributions (red lines). The variance ($\sigma^2$) of the distribution versus the average peak time is shown in Fig. 2d. It is clear that the variance goes linearly with time, confirming the diffusive nature of the phase, and the linear slope defines a peak time diffusion constant $D$. It is easy to show that the coherence time $\tau_c$ is inversely proportional to $D$:
\begin{equation}
\tau_c=\alpha T^2/D,
\end{equation}
where $\alpha$ is a constant dependent on the waveform ($\alpha=(2\pi^2)^{-1}$ for a sine wave).

\subsection{Free energy dissipation suppresses phase diffusion}

As $\gamma$ decreases below $\gamma_c$, more free energy is dissipated. What is the effect of the additional free energy dissipation beyond the onset of oscillation? From the chemical reaction rates, we can compute the free energy dissipation rate \cite{Qian2007}:
\begin{equation}
\dot{W}=\sum_i(J_i^+-J_i^-)\ln{\frac{J_i^+}{J_i^-}}
\label{eq7}
\end{equation}
where $J_i^+$ and $J_i^-$ are the forward and backward fluxes of the $i$th reaction, and free energy is in units of $k_B T$, set to unity here. For the activator-inhibitor and glycolysis models, we calculated the energy dissipation rate using Eq. \ref{eq7}. For systems with continuum stochastic dynamics described by Langevin equations (e.g., the brusselator and the repressilator models), we can obtain the steady-state distribution $P(\vec{x})$ by solving the corresponding Fokker-Planck equation or by direct stochastic simulations (see Fig. S2 in SI for an example). From $P(\vec{x})$, we computed the phase space fluxes and the free energy dissipation rate following \cite{LanTu2013} (see the Methods section and SI for details). For oscillatory systems, the dissipation rate $\dot{W}$ varies in a period $T$. We define $\Delta W\equiv \int_0^T \dot{W} dt$ to characterize the free energy dissipation per period per volume.   

For each of the four models, $\Delta W$ and the dimensionless peak time diffusion constant $D/T$ were computed for different parameter values (reaction rates, protein concentrations) in the oscillatory regime $\gamma<\gamma_c$ and for different volume $V$. As shown in Fig. 3, for all the four models considered, $D/T$ decreases as the energy dissipation $\Delta W$ increases and eventually saturates to a fixed value when $\Delta W\rightarrow \infty$ (i.e., $\gamma=0$). The phase diffusion constants scale inversely with the volume $V$. As shown in the insets of Fig. 3, the scaled $D/T$ (by the volume $V$) collapsed onto a simple curve, which can be approximated by the same simple form in all the four models studied:
\begin{equation}
V\times \frac{D}{T}\approx C+ \frac{W_0}{\Delta W-W_c},
\label{DDW}
\end{equation}
where $W_c$ is the critical free energy, and $W_0$ and $C$ are intensive constants (independent of volume), whose values in different systems (models) are given in the legend of Fig. 3.


\subsection{The free energy sources and experimental evidence}

What is the free energy source driving the biochemical oscillations? For the activator-inhibitor model, the free energy is provided by ATP hydrolysis in the phosphorylation-dephosphorylation (PdP) cycle (see Fig. 1a). Besides the standard free energy $\Delta G_0$ of ATP hydrolysis, the total free energy dissipation per period $\Delta W$ also depends on (and thus can be controlled by) the concentrations of ATP, ADP and the inorganic phosphate $P_i$. These concentrations ([ATP], [ADP], and $[P_i]$) directly affect the biochemical reaction rates in our model and consequently the phase diffusion of the oscillation. In Fig. 4a, we show the phase diffusion constant ($D/T$) versus the dissipation per period ($\Delta W$) for $300$ randomly chosen points in the oscillatory regime of the ($[ATP]$, $[ADP]$, $[P_i]$) space (see Fig. 4b). Remarkably, all the points lie above an envelope curve (the dotted line), which follows Eq.\ref{eq7}. This envelope curve defines the best performance of the biochemical network, i.e., the minimum free energy $\Delta W_m$ needed to achieve a given level of phase coherence. For each choice of the concentrations $([ATP],[ADP],[P_i])$, a functional efficiency $E$ can be defined as the ratio of $\Delta W_m$ and the actual cost $\Delta W$ for the same performance ($D/T$). The efficiency is represented by color in Fig. 4a\&b. We investigated how efficiency depends on the three concentrations. As shown in Fig. 4c, the efficiency $E$ does not simply increase with the ATP concentration; instead it peaks near a particular level of [ATP], at which the phosphorylation and dephosphorylation fluxes are matched. Similarly, $E$ does not have any clear dependence on $[ADP]$ or $[P_i]$ level, it is high near a fixed ratio of $[ADP]/[P_i]$, when the kinetic rates of the phosphorylation and dephosphorylation parts of the PdP cycle are matched.

These predicted dependence of oscillatory behaviors on [ATP], [ADP], and $[P_i]$ concentrations, as shown in Fig. 4, may be tested experimentally by measuring peak-to-peak time variations or equivalently the correlation time for different nucleotide concentrations. As reported in two recent studies \cite{Rust2011,Phong2013}, the oscillatory dynamics of the phosophorylated KaiC protein in a reconstituted circadian clock from cyanobacteria (the Kai system) have been measured in media with different ATP/ADP ratios. We analysed the data according to Eq. 1 and obtained the correlation time ($\tau_c$) and the period ($T$) for different ATP/ADT ratios (see SI and Fig. S3 in SI for details). In Fig. \ref{fig_exp}, we plotted the period and the phase diffusion versus $\ln([ATP]/[ADP])$, which is the entropic contribution to the free energy dissipation. As the ATP/ADP ratio increases, the period changes little. In contrast, the phase diffusion $T/\tau_c\equiv \alpha^{-1} D/T$ decreases significantly and eventually saturates at high ATP/ADP ratios, consistent with the relationship between energy dissipation and phase diffusion discovered here.

\subsection{Analytical results from the noisy Stuart-Landau equation}
To understand the relationship between phase accuracy and energy dissipation, we consider the noisy Stuart-Landau equation for a complex order parameter $Z$:
 \begin{equation}
 \frac{dZ}{dt}=(a+ib)Z-(c+id)|Z|^2Z+\eta_Z,
 \label{STeq}
 \end{equation}
 where $a$, $b$, $c(>0)$, $d$ are real variables, $i=\sqrt{-1}$, and $\eta_Z$ is a complex noise term. For $a>0$, the system starts to oscillate with a mean amplitude $r_s=\sqrt{\frac{a}{c}}$. Eq. (\ref{STeq}) can be decomposed into two Langevin equations for the amplitude $r$ and the phase $\theta$ of $Z=r e^{i\theta}$:
\begin{equation}
\frac{dr}{dt}=ar-cr^3+\eta_r(t) \;\; , \;\; \frac{d\theta}{dt}=b-dr^2+\eta_\theta(t),
\label{eq1}
\end{equation}
where $\eta_r$ and $\eta_\theta$ are the white noises of the amplitude and the phase. For simplicity, we consider the case where $\eta_r$ and $\eta_\theta$ are uncorrelated with constant strength $\Delta_r$ and $\Delta_\theta$ respectively. The average phase velocity is $\omega(r)\equiv \langle d\theta/dt\rangle =b-dr^2$.

It is clear from Eq. \ref{eq1} that detailed balance is broken and the system is dissipative. To compute the free energy dissipation, we first determine the phase-space probability distribution function $P(r,\theta,t)$, which follows the Fokker-Planck equation:
\begin{equation}
\frac{\partial{P}}{\partial{t}}=-\frac{1}{r}\frac{\partial}{\partial r}\big[(ar^2-cr^4)P-\frac{\Delta_r r}{2}\frac{\partial P}{\partial r}\big]-\frac{\partial}{\partial \theta}\big[(b-dr^2)P-\frac{\Delta_\theta}{2}\frac{\partial P}{\partial \theta}\big]\equiv -\frac{1}{r} \frac{\partial (rJ_r)}{\partial r}-\frac{\partial J_\theta}{\partial \theta},
\label{eq2}
\end{equation}
where $J_r$ and $J_\theta$ are the probability density fluxes in phase space. Since $\omega(r)$ does not depend on $\theta$, the steady state probability distribution $P_s(r,\theta)$ only depends on $r$:
\begin{equation}
P_s(r,\theta)=P(r)=A\exp{[-\frac{2(cr^4/4-ar^2/2)}{\Delta_r}]}
\label{eq3}
\end{equation}
where $A=\big[ 2\pi \int{\exp{[-2(cr^4/4-ar^2/2)/\Delta_r]}rdr}\big ]^{-1}$ is the normalization constant. 
From Eq. \ref{eq3}, the flux vanishes in the $r$-direction $J_r=0$. However, there is a finite flux in the $\theta$-direction $J_\theta (r)=\omega(r)P(r)$. 
We compute the system's entropy production rate $\dot{S}$ \cite{Tomede-Oliveira2010,Seifert2005}, from which we obtain the minimum free energy dissipation (see SI for details):
\begin{equation}
\dot{W}=k_B T_{e}\int\int [\frac{J_r^2}{\Delta_r P}+\frac{J_\theta^2}{\Delta_\theta P}]rdrd\theta= k_B T_{e} \frac{\langle \omega^2\rangle}{\Delta_\theta},
\label{eq4}
\end{equation}
where $T_{e}$ is an (effective) temperature of the environment, we set $k_BT_{e}=1$ here. 

The phase diffusion constant is determined by expanding the phase velocity $\omega(r)$ around $r=r_s$, the most probable amplitude from $P(r)$. This leads to  $d\theta/dt = \omega(r_s)+\beta \delta r(t)+\eta_\theta(t)$, with $\beta\equiv\partial\omega(r_s)/\partial r=-2d\sqrt{a/c}$. The period of the oscillation is $T=2\pi/\omega(r_s)$, and the phase fluctuation $\delta \theta\equiv\theta-\omega(r_s)t$ follows diffusion with the diffusion constant given by:
\begin{equation}
D_{\theta}=\frac{\beta^2\Delta_r}{4a^2}+\Delta_\theta.
\label{eq5}
\end{equation}
From Eq. \ref{eq4}\&\ref{eq5}, the relation between phase diffusion and energy dissipation emerges:
\begin{equation}
D_{\theta}=D_0+\frac{\langle \omega^2\rangle T}{\Delta W}\equiv C +\frac{W_0}{\Delta W-W_c},
\label{eq6}
\end{equation}
where $W_c=0$ because $J_r=0$, something that is not generally valid (see SI and Fig. S4 in SI for a more general case of the noisy Stuart-Landau equation). The two constants, $C=D_0=\frac{\beta^2\Delta_r}{4a^2}$ and $W_0=\langle \omega^2\rangle T$, depend on the details of the system. 

Eq.\ref{eq6} has the same form as Eq. \ref{DDW} obtained empirically from studying different biochemical networks. Analysis of the noisy Stuart-Landau equation clearly shows that free energy dissipation is used to suppress phase diffusion to increase the coherence of the oscillation. Though parameters in this relation may depend on the details of the system, the inverse dependence of phase diffusion on energy dissipation appears to be universal.

\section{discussion}

Oscillations are critical for many biological functions that require accurate time control, such as circadian clock, cell cycle, and development. However, biological systems are inherently noisy. The phase of a noisy oscillator fluctuates (diffuses) without bound and eventually destroys the coherence (accuracy) of the oscillation.
Specifically, the number of periods $N_c$ in which the oscillation maintains its phase coherence is given by $N_c=\tau_c/T =\alpha T/D$, which decreases with the phase diffusion constant. Here, our study shows that free energy dissipation can be used to reduce phase diffusion and thus prolong the coherence of the oscillation. A general relationship between the phase diffusion constant and the minimum free energy cost, as given in Eq. \ref{DDW}, holds true for all the oscillatory systems we studied here. The amplitude fluctuations also decrease with free energy dissipation (see Fig. S5 in SI for details), as fluctuations in phase and amplitude are coupled in realistic systems. Our study thus establishes a cost-performance tradeoff for noisy biochemical oscillations.

How do biological systems use their free energy sources (e.g., ATP) to enhance the accuracy of the biochemical oscillations? As illustrated in Fig. 5a, a biochemical oscillation can be considered as a clock, which goes through a series of time-ordered chemical states (green dots) during each period. These chemical states are characterized by the conformational and chemical modification (e.g., phosphorylation) states of the key proteins or protein complexes in the system. The forward transition from one state to the next is coupled to a PdP cycle (blue arrowed circle) driven by hydrolysis of one ATP molecule. For each forward step, the reverse transition introduces a large error in the clock. The system suppresses these backward transitions by utilizing the ATP hydrolysis free energy. However, this is just one half of the story. Even in the absence of the reverse transition, the time duration between two consecutive states is highly variable due to the stochastic nature (Poisson process) of the chemical transitions. A general strategy of increasing accuracy is averaging \cite{bergpurcell}. In the case of biochemical oscillations, each period may consist of multiple steps, each powered by at least one ATP molecule. As a result of averaging, the error in the period should go down as the number of steps increases. Specifically, we expect that the variance of the period $\sigma_T^2(=D/T)$ should be inversely proportional to the total number of ATP hydrolyzed $ N_{ATP} \propto T/\tau_{cyc}$ in each period $T$, where $\tau_{cyc}$ is the average PdP cycle time, which is essentially the ATP turnover time. Consequently, the number of coherent period $N_c=\alpha T/D$ should be proportional to the number of ATP hydrolyzed in each period. We checked this prediction by varying the kinetic rates in the PdP cycle to change $\tau_{cyc}$ (see Methods section for details). In Fig. 5b, it is shown that the accuracy of the oscillation (clock), as measured by $N_c$, is enhanced by the number of ATP molecules hydrolyzed in each period. This result reveals a general strategy for oscillatory biochemical networks to enhance their phase coherence by coupling to multiple energy consuming cycles in each period. Interestingly, approximately $15$ ATP molecules are consumed per KaiC molecule per period in the circadian clock of cyanobacteria \cite{Terauchi2007}.

Biological systems need to function robustly against variations in its underlying biochemical parameters (rates, concentrations) \cite{BarkaiLeibler1997,MaTrusina2009}. For oscillatory networks, the free energy dissipation needs to reach a critical value ($W_c$) to drive the system to oscillate. We showed here that additional free energy cost in excess of $W_c$ is needed to make the oscillation more accurate, as demonstrated explicitly in Eq. \ref{DDW}. In addition to this accuracy-energy tradeoff, we found that larger energy dissipation can also enhance the system's robustness against its parameter variations. Take the activator-inhibitor model, for example: the concentrations of enzyme E ($E_T$) and phosphatase K ($E_K$) may vary from cell to cell.
We search for the existence of oscillation in the ($E_T$, $K_T$) space for different values of $\gamma$. Robustness is defined as the area of the parameter space where oscillation exists. As shown in Fig. S6 in the SI, the robustness increases as the system becomes more irreversible, i.e., when more free energy dissipation is dissipated. This suggests a possible general tradeoff between the functional robustness and energy dissipation in biological networks.
\section{Acknowledgement}
We thank Dr. Michael Rust for sharing the experimental data in Ref. \cite{Rust2011}\&\cite{Phong2013} with us. This work is partly supported by a NIH grant (R01GM081747 to YT).



\section{Methods}
{\bf Simulation Methods.} The Gillespie algorithm \cite{Gillespie1977} is used for the stochastic simulations of the reaction kinetics. 
For given kinetic rates and the volume $V$, we simulated $1000$ trajectories starting with the same initial condition. For the $j$th trajectory, we obtained its $i$th peak time $t_{ij}$ from the trajectory $x_j(t)$ after smoothing (smooth function in MATLAB was used). The peak positions for two trajectories are shown in Fig. 2a. 
For all the trajectories, we computed the mean of their $i$th peak time $m_i=\sum_j t_{ij}/N$, and variance $\sigma^2_i=\sum_j(t_{ij}-m_i)^2/(N-1)$, where $N$ is the total number of trajectories. The average period $T$ is given by $T=m_i/i$. Asymptotically, $\sigma_i^2$ depends linearly on $m_i$ (Fig. 2d), and the slope of this linear dependence is the peak time diffusion constant $D$, which has the dimension of time. The phase diffusion constant $D_{\phi}$ is linearly proportional to $D$: $D_{\phi}=(2\pi)^2D/T$. For the repressilator and the brusselator models, we simulated the stochastic kinetic equations to a sufficiently long time ($10000$ periods) to obtain the time-averaged distribution $P(\vec{x})$, where $\vec{x}$ represents the phase space. We used $P(\vec{x})$ to compute free energy dissipation.  

{\bf Random Sampling} in the $([ATP],[ADP],[P_i])$ space is performed (in log scale) in the region $\log_{10}\frac{[ATP]}{[ATP]_0}\in [2,5],\log_{10}\frac{[ADP]}{[ADP]_0}\in [-3,1],\log_{10}\frac{[P_i]}{[P_i]_0}\in [-3,1]$ by using Latin hypercube sampling (the lhsdesign function in MATLAB). The reference concentrations $[ATP]_0$, $[ADP]_0$, and
$[P_i]_0$ are set to unity and their actual values are absorbed into the baseline reaction rates $a_{1,0}$, $f_{-1,0}$ and $f_{-2,0}$, which are given in the legend of Fig. 4. 

{\bf ATP consumption.} 
In the activator-inhibitor model, the ATP consumption rate is $R_{ATP}=V(J_p^+-J_p^-)$, where $J_p^+$ and $J_p^-$ are the fluxes for the $E\rightarrow E_p$ and $E_p\rightarrow E$ reactions, respectively. We varied the overall reaction kinetics, e.g., $\tau_{cyc}$ and the ATP consumption rate, by introducing a timescale factor $B$ for all four rates $d_1=d_2=Bd,f_{1}=f_{2}=Bf$, where $d=15,f=15$ are the original values used in this paper (see SI). By changing the rates this way, the free energy release of ATP hydrolysis $\Delta G=-\ln{\gamma}=\ln(a_1f_1a_2f_2/(d_1f_{-1}d_2f_{-2}))$
is unchanged. We varied $B\in [0.2,2]$, and computed the total number of ATP consumed per period $N_{ATP}\equiv \int_0^T R_{ATP}dt$ and $N_c$ for Fig. 5b.

\section{Acknowledgement}
We thank Dr. Michael Rust for sharing the experimental data in Ref. \cite{Rust2011}\&\cite{Phong2013} with us. This work is partly supported by a NIH grant (R01GM081747 to YT).

\bibliographystyle{naturemag}
\bibliography{references}

\begin{FPfigure}
\centering
\includegraphics[width=13cm]{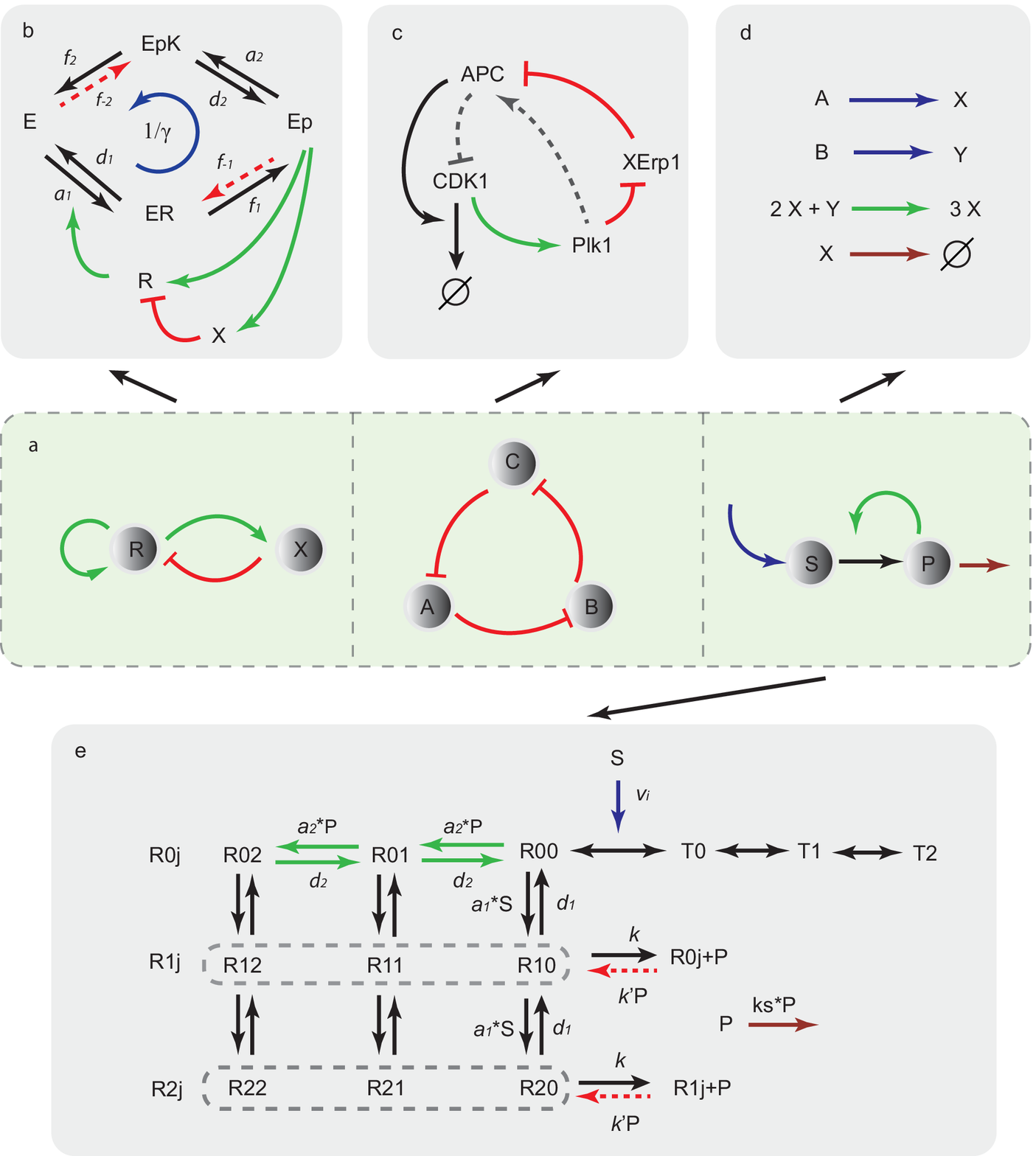}
\caption{Different network motifs and the corresponding biochemical oscillatory systems. (a) Illustrations of three network motifs for oscillation: activator-inhibitor, repressilator, and substrate-depletion.  (b) The activator-inhibitor model with a phosphorylation-dephosphorylation (PdP) cycle. R and K catalyse two opposing  reactions $E\leftrightarrow E_p$ (phosphorylation and dephosphorylation) through different intermediate complexes $ER$ and $E_pK$. $E_p$ activates both $R$ (activator) and $X$ (inhibitor). $X$ inhibits $R$ by enhancing its degradation. Parameter $\gamma=d_1f_{-1}d_2f_{-2}/(a_1f_1a_2f_2)$ is introduced to characterize the reversibility of the system. (c) The ``repressilator" model of cell cycle in eukaryotic cells. In the simplified network, CDK1 activates Plk1, Plk1 activates APC, and APC degrades CDK1 (dashed line), forming the mutually activing/inhibiting loop. Other intermediates are ignored here. (d) The brusselator model with detailed reactions. A and B are constant sources. (e) The glycolysis network. The allosteric enzyme's protomer has two states, R (binding with P) and T (unbinding with P), and only R has the catalysis activity. Each $R_{i,j}$, with $i=1,2,\cdots,n_i$ and $j=1,2,\cdots,n_j$ represent the number of $S$ and $P$ bound to $R$, here we used $n_i=n_j=2$. Each $R_{i,j}$ can undergo reactions of $R_{i,j}+S\leftrightarrow R_{i,j+1}\leftrightarrow R_{i,j}+P$. Detailed descriptions and rate values are given in SI.}
\label{fig1}
\end{FPfigure}

\begin{figure}
\centering
\includegraphics[width=15cm]{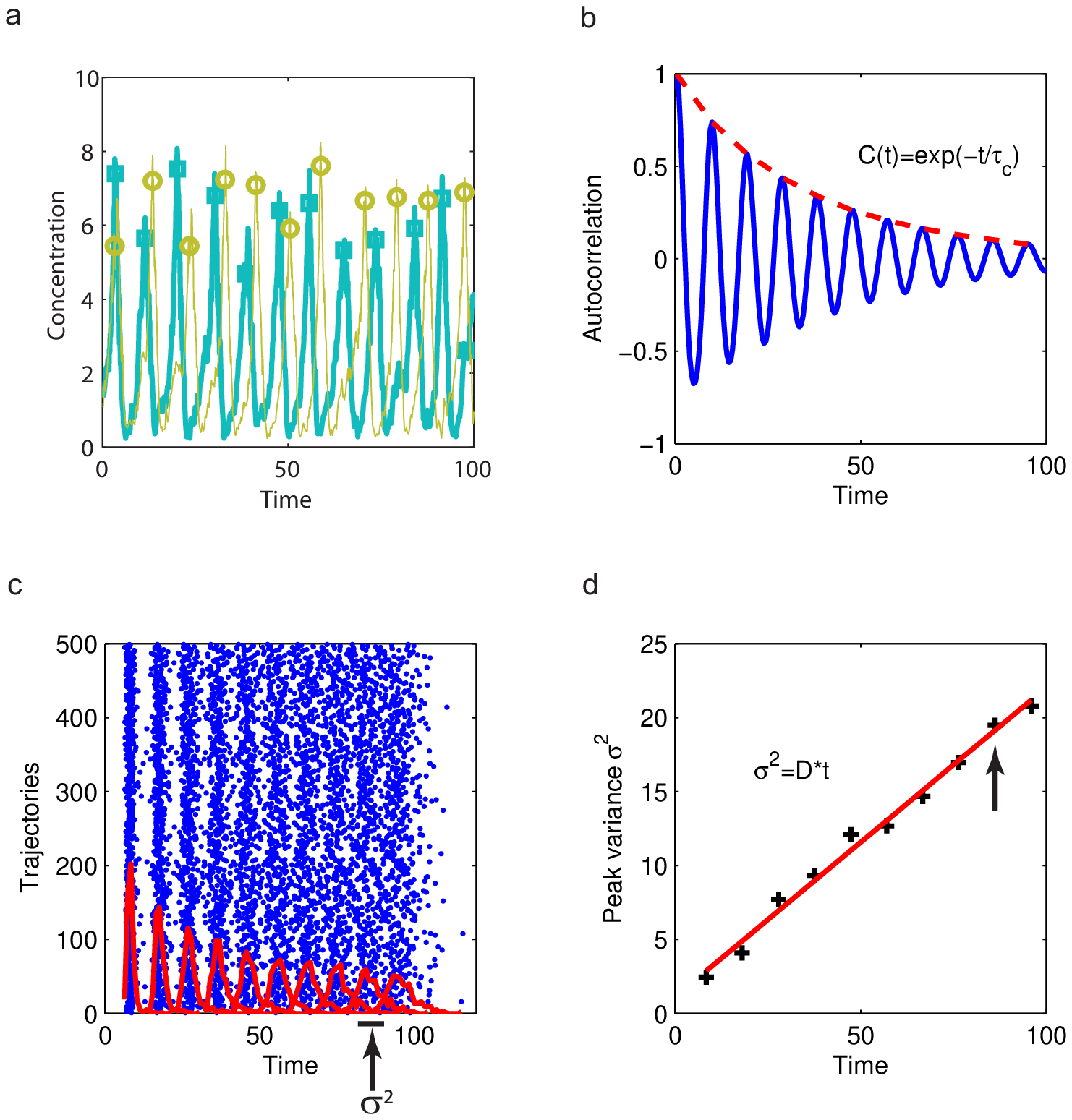}
\caption{Correlation and phase diffusion in the activator-inhibitor model with $V=50, \gamma=10^{-5}$. (a) Two noisy oscillation trajectories, with the peaks labeled by circles and squares. (b) Auto-correlation function (defined in Eq. 1) of the inhibitor $X$. $C(t)$ decays exponentially with correlation time $\tau_c=37.7$. (c) Raster plot of the peak times for $500$ different trajectories starting with the same initial condition. The distributions of the peak times for each consecutive peaks are shown by red lines. The peak time variance $\sigma^2$ is shown. (d) Peak time variance $\sigma^2$ goes linearly with the average peak time, with the linear coefficient defined as the peak time diffusion constant. Here, the diffusion constant $D=0.2$ and $\alpha\equiv \tau_c D/T^2\approx0.07$. }
\end{figure}

\begin{figure}
\centering
\includegraphics[width=15cm]{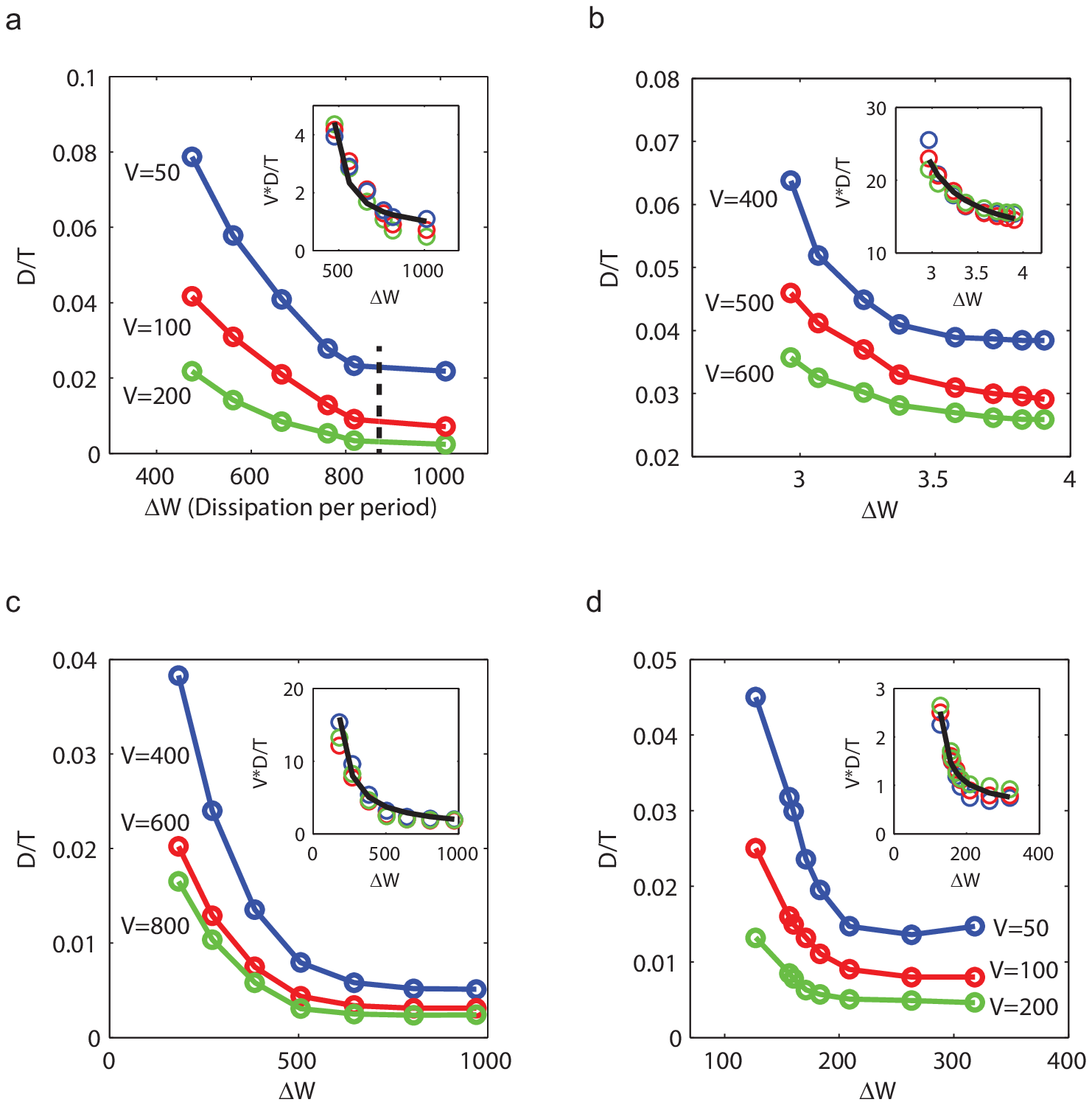}
\caption{Relation between the dimensionless diffusion constant ($D/T$) and free energy dissipation per period per volume ($\Delta W$, in units of $k_BT$) for the four oscillatory systems. Detailed descriptions of the models and parameters can be found in SI. The relationships for different volumes collapse onto the same curve when the peak time diffusion constant is scaled by $V$, as shown in the insets. The black dashed line in the activator-inhibitor model indicates the value of $\Delta W$ if we assume that hydrolysis of one ATP molecule provides $\approx 12k_BT$ energy, which corresponds to $\gamma\approx 10^{-5.2}$. All the data can be well fitted with Eq. 4: $V\times D/T=C+W_0/(W-W_c)$ (lines in insets), where $W_c$ is determined from the critical value $\gamma_c$, $W_0$ and $C$ are from fitting. The parameters are: (a) activator-inhibitor, $W_c=360.4,W_0=447.3\pm 55.8,C=0.28\pm 0.16$; (b) repressilator, $W_c=1.9,W_0=17.7\pm 3.9,C=5.5\pm2.5$; (c) brusselator, $W_c=93.1,W_0=1135\pm 142,C=0.27\pm 0.11$; (d) glycolysis, $W_c=67.4,W_0=135.8\pm3.4,C=0.025\pm0.019$.}
\end{figure}

\begin{figure}
\centering
\includegraphics[width=16cm]{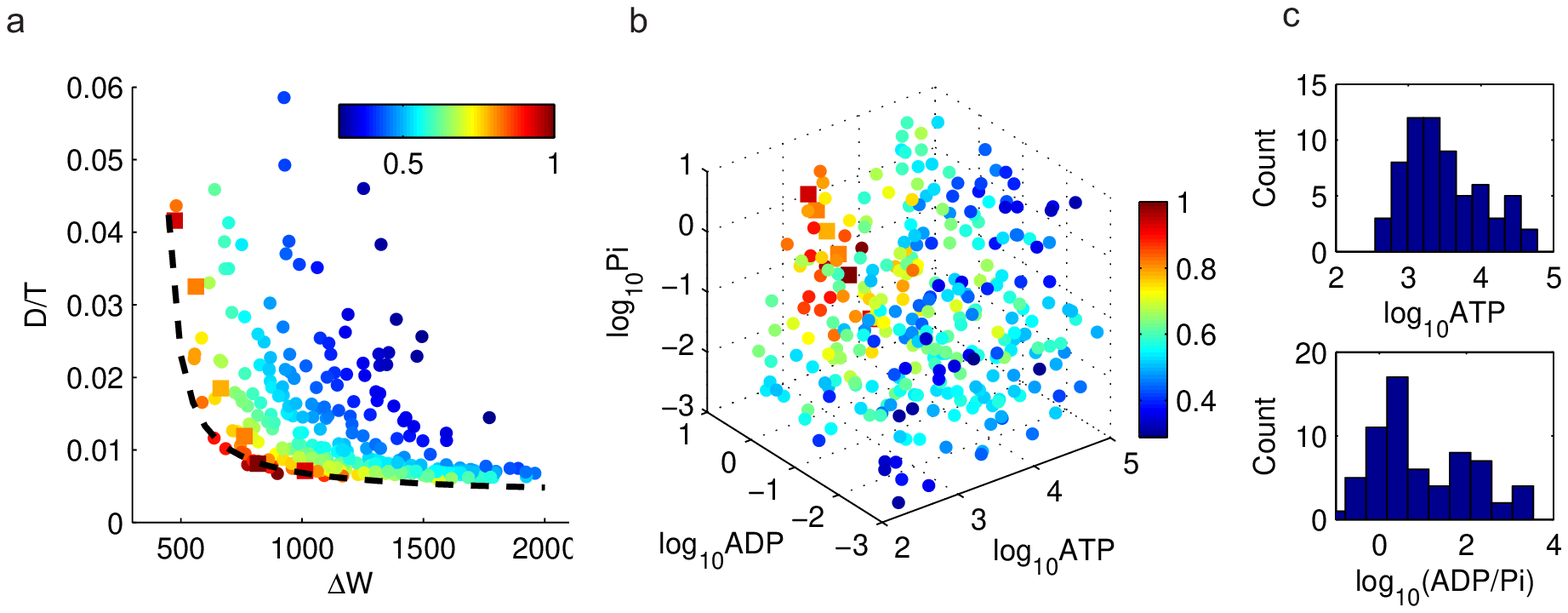}
\caption{The dependence of phase diffusion on the ATP, ADP, and $P_i$ concentrations. We studied the activator-inhibitor model with $300$ randomly chosen parameters of dimensionless [ATP],[ADP] and $[P_i]$ (see Methods). The affected kinetic rates are $a_1=a_{1,0}[ATP],f_{-1}=f_{-1,0}[ADP],f_{-2}=f_{-2,0}[P_i]$ with $a_{1,0}=0.1, f_{-1,0}=f_{-2,0}=1$. We chose $V=100$. (a) $D/T$ versus $\Delta W$ for the $300$ different parameter choices. All the points lie above an envelope curve, which follows Eq. 4 with $D/T=1.94/(\Delta W-400)+0.0036$. Points in square indicate the points shown in Fig. 3. For a given value of $D/T$, the corresponding minimal energy dissipation $\Delta W_{min}$ is computed according to the fitted envelope curve. The efficiency is defined as $E\equiv \Delta W_{min}/\Delta W$. Colors of the points indicate the efficiency. (b) Distribution of the $300$ randomly sampled points in the parameter space $([ATP],[ADP],[P_i])$. Colors of the points indicate the efficiency as in (a). Points with high efficiency are clustered. (c) Distribution of [ATP], and [ADP]/[Pi] for parameter choices with high efficiency $E\geq 0.75$. The most probable parameter values for high efficiency are $[ATP]\approx 10^3,[ADP]\approx [P_i]$, which corresponds to $a_1=a_2,f_{-1}=f_{-2}$ in the kinetic equations. This result indicates that high efficiency is achieved when the kinetic rates in the two halves of the PdP cycle (phosphorylation and dephosphorylation) are matched.}
\end{figure}

\begin{figure}
\centering
\includegraphics{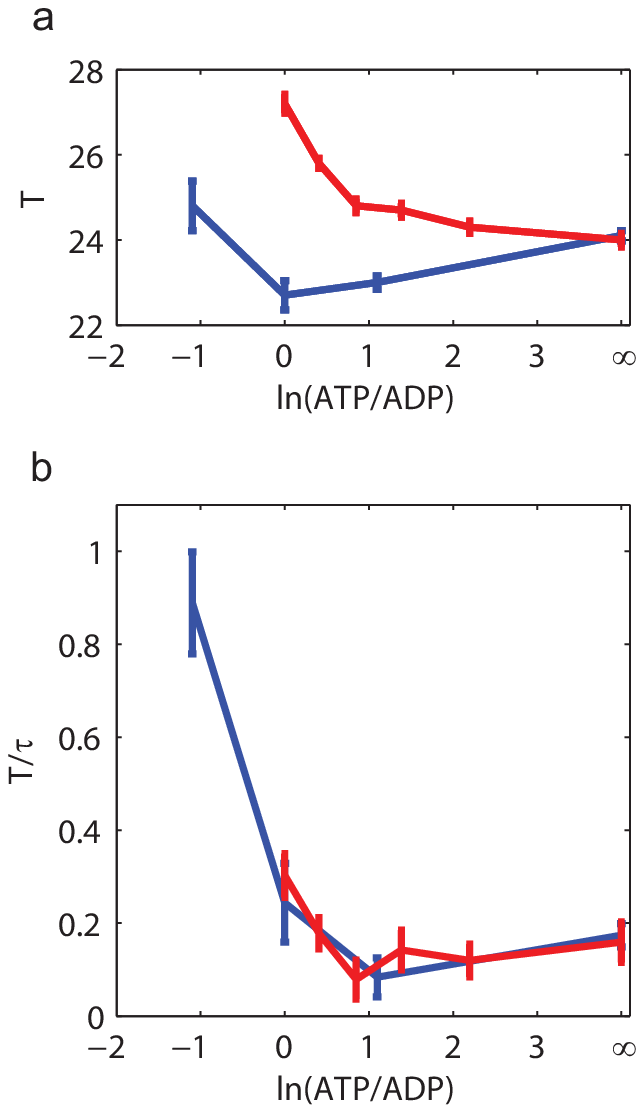}
\caption{Experimental evidence from Ref.\cite{Phong2013} (blue curve) and Ref.\cite{Rust2011} (Red curve). The two experiments measured the oscillation of KaiC phosphorylation in vitro in media with different ATP/ADP ratios. The autocorrelation functions were calculated from the original data and fitted by a exponential decay cosine function $A\cos{(2\pi t/T)}e^{-t/\tau_c}$, where $T$ is the period, and $\tau_c$ is the correlation time. $\ln(ATP/ADP)$ represents the entropic contribution to the free energy. (a) The period $T$ is robust against changes in the ATP/ADP ratio. (b) $T/\tau_c\equiv \alpha^{-1} D/T$ decreases with $\ln(ATP/ADP)$ and eventually saturates at large ATP/ADP ratio, consistent with our theoretical prediction.}
\label{fig_exp}
\end{figure}

\begin{figure}
\centering
\includegraphics[width=15cm]{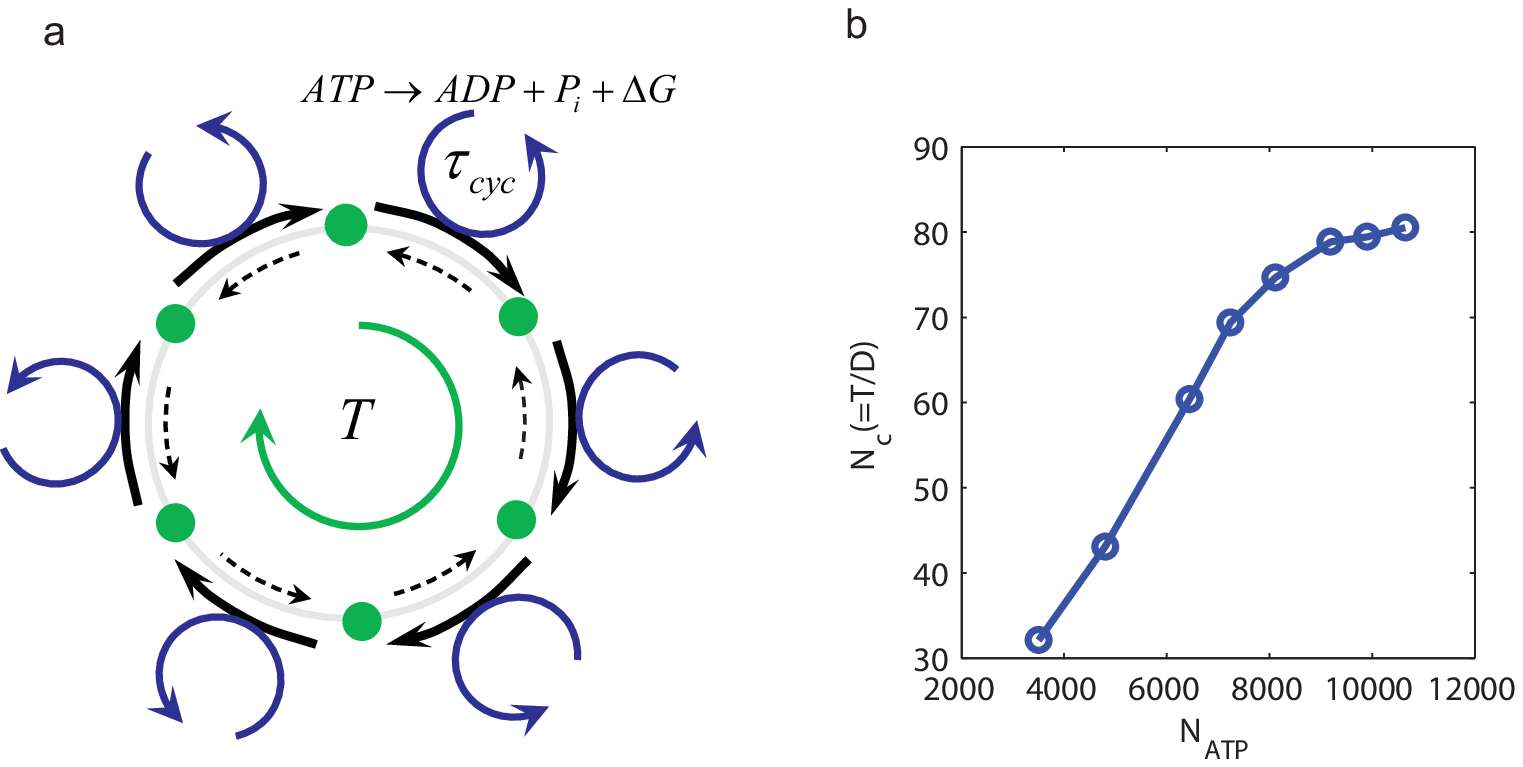}
\caption{Oscillation coherence increases with the number of ATP hydrolyzed per period. (a) Illustration of a biochemical oscillation as a clock in phase space. The intermediate states (green dots) are represented as the "hour ticks" of the clock. The transition from one tick to the next is coupled with a ATP hydrolysis cycle. The free energy release $\Delta G$ from the hydrolysis cycle powers the forward transition (thick solid arrow) and/or suppresses the backward transition (thin dotted arrow). The number of ATP consumed per enzyme molecule in each period $T$ is given by $T/\tau_{cyc}$, where $\tau_{cyc}$ is the average cycle time. (b) The accuracy of the oscillation, characterized by  the number of correlated (coherent) periods $N_c$, increases linearly with the total number of ATP consumed per period $ N_{ATP}$ before saturating at very high $N_{ATP}$. We varied $ N_{ATP}$ by changing the cycle time (see Methods for details), we used $V=100$ here. }
\end{figure}

\end{document}


\begin{centering}
{\Large \bf Supplementary Information for}\\
\vspace{1.0pc}
{\Large \bf ``The free energy cost of accurate biochemical oscillations" by Cao et al.}\\
\end{centering}

\vspace{1.5pc}

\section{descriptions of the four models}

Here, we describe the mathematical details of the four models of biochemical oscillations studied in this paper. For the activator-inhibitor and the glycolysis model, we only give the ordinary differential equations (ODE's) to describe the deterministic part of the chemical reactions. The actual simulations of the stochastic reactions were done using Gillespie algorithm (see Methods in the main text). For the repressilator and brusselator models, we use the chemical master equation (or Langevin equation with Poisson noise)
and solve the corresponding Fokker-Planck equation:
\begin{equation}
\frac{\partial P(\vec{x},t)}{\partial t}=-\nabla(\mathbf{F}P-\mathbf{D}\nabla P)=-\nabla J,
\label{FPE}
\end{equation}
where $\mathbf{F}$ is the force vector, and $\mathbf{D}$ is the noise matrix. $J$ is the flux vector for each direction.

In all our models, the units of the parameters are composed of concentration ($c$, arbitrary), time ($t$, arbitrary) and volume ($V$, arbitrary). The molecule number unit $c\times V$ represents the real counts of a given molecule in the system. For example, if the concentration of enzyme $E$ is $E_T=10(c)$, and the volume is $V=100(V)$, then the total number of enzyme $E$ in our system is $E_TV=1000$.

\subsection{Activator-inhibitor model}
The main components of the model are the activator R and its inhibitor X. R and X are linked in a feedback loop through a phosphorylation-dephosphorylation (PdP) cycle (main text Fig. 1b). R activates the synthesis of both R and X through phosphorylated enzyme E, thus forms a positive feedback; at the same time, X degrades R, thus forms a negative feedback. The parameter $\gamma=d_1d_2f_{-1}f_{-2}/(a_1a_2f_1f_2)$ is introduced to distinguish wether the system is in equilibrium ($\gamma=1$) or non-equilibrium ($0<\gamma<1$). The kinetics is described by:
\begin{equation}\begin{split}
&\frac{d[R]}{dt}=k_0[E_p]+k_1S-k_2[X][R]\\
&\frac{d[X]}{dt}=k_3[E_p]-k_4[X]\\
&\frac{d[E]}{dt}=f_2[E_pK]+d_1[ER]-a_1[E]([R]-[ER])-f_{-2}[E][K]\\
&\frac{d[ER]}{dt}=a_1[E]([R]-[ER])+f_{-1}[E_p]([R]-[ER])-(f_1+d_1)[ER]\\
&\frac{d[E_p]}{dt}=f_1[ER]+d_2[E_pK]-a_2[E_p][K]-f_{-1}[E_p]([R]-[ER])\\
&\frac{d[E_pK]}{dt}=a_2[E_p][K]+f_{-2}[E][K]-f_2[E_pK]-d_2[E_pK]\\
\end{split}
\end{equation}
with two mass conservation constraints: $[E]+[E_p]+[ER]+[E_pK]=E_T, [E_pK]+[K]=K_T$, where $E_T$ and $K_T$ are the total concentrations of enzyme E and phosphatase K. Each term in the equations represents one of the reactions in the main text (see Fig. 1b). We take symmetric parameters: $k_0=k_1=k_3=1(t^{-1}),k_2=1(c^{-1}t^{-1}), k_4=0.5(t^{-1}),S=0.4(c),K_T=1(c),E_T=10(c),a_1=a_2=100(c^{-1}t^{-1}),f_1=f_2=d_1=d_2=15(t^{-1}),f_{-1}=f_{-2}=\sqrt{\gamma}a_1f_1/d_1(c^{-1}t^{-1})$. The oscillation onset point is at $\gamma_c=2\times10^{-3}$.  Notice that the PdP cycle's reaction rates are much larger than the reactions of synthesis and degradation of R and X, so the total R is almost unchanged in the time scale of the PdP cycles.
\subsection{Repressilator}
We use the simplified cell cycle model in\cite{FerrellTsai2011}, where CDK activates Plk1, and Plk1 activates APC, which inhibits CDK in return. The (deterministic) negative feedback loop kinetics are governed by the following ODE's, with CDK, Plk1, APC concentrations represented by $x,y,z$, respectively:
\begin{equation}\begin{split}
&\frac{dx}{dt}=\alpha_1-\beta_1\frac{z^{n_1}}{K_1^{n_1}+z^{n_1}}=f_x-d_x\\
&\frac{dy}{dt}=\alpha_2(1-y)\frac{x^{n_2}}{K_2^{n_2}+x^{n_2}}-\beta_2y=f_y-d_y\\
&\frac{dz}{dt}=\alpha_3(1-z)\frac{y^{n_3}}{K_3^{n_3}+y^{n_3}}-\beta_3z=f_z-d_z\\
\end{split}
\end{equation}
where $f_i,d_i$ are the synthesis and decay rates of each component. We chose $\alpha_1=0.1(ct^{-1}),\alpha_2=3(t^{-1}),\beta_1=3(ct^{-1}),\beta_2=1(t^{-1}),\beta_3=1(t^{-1}),K_1=0.5(c),K_2=0.5(c),K_3=0.5(c),n_1=8,n_2=8,n_3=8$, and $\alpha_3(t^{-1})$ is taken as the control parameter ranges from 1.0 to 3.0. The oscillation onset point is $\alpha_3=0.8$. 

The full stochastic dynamics is described by the Fokker-Planck equation (Eq. \ref{FPE}) with the force vector and noise matrix given by:
$$\mathbf{F}=[f_x-d_x,f_y-d_y,f_z-d_z],\;\;\;
\mathbf{D}=\frac{1}{2V}diag[f_x+d_x,f_y+d_y,f_z+d_z]$$.
\subsection{Brusselator}
The deterministic equation of brusselator is
\begin{equation}\begin{split}
&\frac{dx}{dt}=a-x+x^2y,\\
&\frac{dy}{dt}=b-x^2y.\\
\end{split}
\end{equation}
The Fokker-Planck equation was derived from chemical master equations (CME) in \cite{QianSaffarian2002}, which gave:
\begin{subequations}
\begin{equation}
\mathbf{F}=\left(\begin{array}{c}a-x+x^2y\\b-x^2y\\
\end{array}\right)+\frac{1}{2V}\left(\begin{array}{c}-1/2-2xy+x^2/2\\2xy-x^2/2\end{array}\right)
\end{equation}
\begin{equation}
\mathbf{D}=\frac{1}{2V}\left(\begin{array}{cc}a+x+x^2y&-x^2y\\-x^2y&b+x^2y\end{array}\right)
\end{equation}
\end{subequations}
where $V$ is the volume of the system. We used $b=0.4$, and varied $a\in[0.12,0.18]$ in our study. Fig. S1 gives two examples of probability distribution $P(\vec{x},t)$ and fluxes $J$ in the brusselator model.
\subsection{Glycolysis oscillation}
The glycolysis model in our study is taken from \cite{GoldbeterLefever1972}, but we have introduced finite reverse reaction rates for the catalysis processes to study the free energy dissipation in glycolysis. The enzyme PFK are composed of $n$ protomers, and undergoes allosteric regulation by its product P. Each protomer exists two states, R, which has catalytic activity for converting substrate $S$ to $P$, and T, which is inactive. Assuming a quasi-equilibrium of the allosteric states of PFK \cite{GoldbeterLefever1972}, the dynamics of $S$ and $P$ can be written as:
\begin{equation}\begin{split}
&\frac{dS}{dt}=v_i-\frac{nD(1+\alpha)^{n-1}(1+\theta)^n(k\alpha-k'P)}{L+(1+\alpha)^n(1+\theta)^n}\\
&\frac{dP}{dt}=\frac{nD(1+\alpha)^{n-1}(1+\theta)^n(k\alpha-k'P)}{L+(1+\alpha)^n(1+\theta)^n}-k_sP\\
\end{split}
\end{equation}
where $\alpha=(a_1S+k'P)/(k+d_1),\theta=a_2P/d_2$. Parameters were chosen as $D=500(c), n=2, a_1=a_2=10(c^{-1}t^{-1}),d_1=d_2=10(t^{-1}), k=1(t^{-1}), v_i=0.2(ct^{-1}), k_s=0.1(t^{-1}),L=7.5\times10^6$. $k'(c^{-1}t^{-1})$ is the reverse reaction rate of P to S. The oscillation onset point is $k'=4\times10^{-1}$. Stochastic simulations were performed for 4 reactions: synthesis of S, degradation of P, catalysis of S to P, reverse reaction of P to S. Here, we combined all the enzymatic reactions into two reactions since the transitions between different allosteric states of the enzyme are much faster than the slow reactions of substrate injection ($v_i$) and product removal ($k_s$).

The dissipation of the system can be directly calculated by summing up the dissipation of all the enzymatic reactions:
\begin{equation}
\Delta W(t)=\frac{nD(1+\alpha)^{n-1}(1+\theta)^n(k\alpha-k'P)}{L+(1+\alpha)^n(1+\theta)^n}\times \log{\frac{ka_1S}{k'd_1P}}=N_S\times\Delta G,
\end{equation}
 where $N_S$ quantifies how many molecules of S are catalysed to P. 

\section{energy dissipation determined from solving the Fokker-Planck equation}
Consider a general Fokker-Planck equation
\begin{equation}
\frac{\partial P(\vec{x},t)}{\partial t}=-\nabla(\mathbf{F}P-\mathbf{D}\nabla P)=-\nabla J
\label{ap_eq1}
\end{equation}
where $\mathbf{F}$ is the force vector, and $\mathbf{D}$ is the noise matrix. $J$ is the flux vector for each direction. The system's entropy is
\begin{equation}
S(t)=-\int P(\vec{x},t)\ln{P(\vec{x},t)}d\vec{x}
\label{ap_eq2}
\end{equation}
 The entropy production rate is\cite{Tomede-Oliveira2010}:
\begin{equation}
\frac{dS(t)}{dt}=-\int[\ln{P(\vec{x},t)}+1]\frac{\partial P(\vec{x},t)}{\partial t}d\vec{x}=\int[\ln{P(\vec{x},t)+1}]\nabla Jd\vec{x}
\label{ap_eq3}
\end{equation}
Integrated by parts:
\begin{equation}
\frac{dS(t)}{dt}=-\int J^T\nabla\ln{P(\vec{x},t)}d\vec{x}
\label{ap_eq4}
\end{equation}
where $J^T$ is the transposition of $J$. By definition $J=\mathbf{F}P-\mathbf{D}\nabla P$, we have
\begin{equation}
J^T\nabla \ln{P}=J^T\mathbf{D}^{-1}\mathbf{F}-\frac{J^T\mathbf{D}^{-1}J}{P}
\label{ap_eq5}
\end{equation}
Finally
\begin{equation}
\frac{dS(t)}{dt}=-\int J^T\mathbf{D}^{-1}\mathbf{F}d\vec{x}+\int\frac{J^T\mathbf{D}^{-1}J}{P}d\vec{x}
\label{ap_eq6}
\end{equation}
The second term is the free energy dissipation rate (also called entropy production rate\cite{Ge2010}) in unit of $k_BT$.

\section{onset of oscillation}
Systems at the onset of oscillation is dissipative. In Fig. S2, we show that at the onset of oscillation, i.e., when amplitude is zero, the free energy dissipation is finite positive. The free energy dissipation at onset defines $W_c$ that we used in fitting main text Fig3.

\section{Experimental data analysis}
The experimental data were obtained from Ref\cite{Phong2013} and Ref\cite{Rust2011}. The data were processed to calculate the autocorrelation function and fitted the autocorrelation function with exponentially decay function $A\cos(2\pi t/T)e^{-t/\tau}$, from which the period $T$ and correlation time $\tau$ could be obtained. See Fig\ref{figS_exp}.

\section{amplitude fluctuation and phase diffusion in the Stuart-Landau equation}

We first derive the amplitude and phase variance in the simplified Langevin equations from the main text (Eq. 6). The deviation in \(r\) can be defined as
\begin{equation}
\delta r=r-r_s,
\end{equation}
where $r_s=\sqrt{a/c}$. Perturbation $\delta r$  near $r_s$ follows the equation (in first order approximation):
\begin{equation}
\frac{d(\delta r)}{dt}=-2a\delta r +\eta_r(t).
\end{equation}
Following \cite{Kampen2007}, as $t\rightarrow\infty$, we obtain the amplitude fluctuation:
\begin{equation}
\langle \delta r^2(\infty)\rangle=\frac{\Delta_r}{4a},
\end{equation}
which only depends on $\Delta_r$.

The phase is $\theta(t)=\int_0^t\omega(r(\tau))d\tau$. With the expanding expression $\omega(r(\tau))=\omega(r_s)+\beta\delta r(\tau)+\eta_\theta(\tau)$, we have the phase variance:
\begin{equation}\begin{split}
\langle\theta^2\rangle-\langle\theta\rangle^2
&= \int_0^t\int_0^t(\langle\omega(\tau_1)\omega(\tau_2)\rangle-\langle\omega(\tau_1)\rangle\langle\omega(\tau_2)\rangle)d\tau_1d\tau_2\\
&=\beta^2\int_0^t\int_0^t[\langle\delta r(\tau_1)\delta r(\tau_2)\rangle-\langle\delta r(\tau_1)\rangle\langle \delta r(\tau_2)\rangle] d\tau_1d\tau_2+\Delta_\theta t.
\end{split}\label{ap_eq12}\end{equation}
Following \cite{Kampen2007}, as  $t\rightarrow\infty$, we have:
\begin{equation}\langle\theta^2\rangle-\langle\theta\rangle^2=
(\beta^2\frac{\Delta_r}{4a^2}+\Delta_\theta)t\equiv D_\theta t.
\label{ap_eq14}
\end{equation}
Clearly, the phase diffusion depends on both $\Delta_\theta$ and $\Delta_r$.

Next, we show a general case of the Stuart-Landau equation described by the Langevin equations for the two real variables $x$ and $y$: $Z=x+iy$.
\begin{equation}\begin{split}
&\frac{dx}{dt}=(ax-by)-(cx-dy)(x^2+y^2)+\sqrt{\Delta_1}\eta_1(t)\\
&\frac{dy}{dt}=(bx+ay)-(cy+dx)(x^2+y^2)+\sqrt{\Delta_2}\eta_2(t)\\
\end{split}
\end{equation}
We studied these two equations numerically. We found that if $\Delta_1\neq\Delta_2$, then $W_c\neq0$. In Fig. S4, we show a case with fixed $\Delta_1=0.1$ while varying $\Delta_2\in[0.05,0.2]$. We found that the onset, which corresponds to $D\rightarrow \infty$, occurs at a finite energy dissipation $W_c\neq0$ (Fig. S4a). However, the general relationship of $D/T=W_0/(\Delta W-W_c)+C$, with $W_c\ne 0$ from Fig. S4a, holds true (Fig. S4b).

\section{amplitude fluctuation}
The amplitude fluctuation can be defined as the dispersion of the stochastic trajectories departing from the deterministic trajectory in phase-space:
\begin{equation}
d^2=\int\min{(\vec{x}-\vec{x}_d)^2}P(\vec{x})d\vec{x}
\end{equation}
where $\vec{x}_d$ are the points on the deterministic trajectory. In the four models we studied, the amplitude fluctuations decrease with energy dissipation and scale with $1/\sqrt{V}$, as shown in Fig. S5.

\section{robustness and energy dissipation}
For the activator-inhibitor model, the total number of enzyme E and phosphatase K may vary in real systems (e.g., from cell to cell). Here we search the parameter space of $(E_T,K_T)$ in the region $E_T\in[0,1000]$ and $K_T\in[0,2]$, and check wether the system oscillates (with amplitude larger than 0.1) for different values of $\gamma$. Robustness is defined as the area in the parameter space where oscillation exists. We found robustness increases as the system becomes more irreversible or equivalently dissipates more free energy, as shown in Fig. S6.

\bibliographystyle{naturemag}
\bibliography{references}

\newpage

\begin{figure}
\centering
\includegraphics{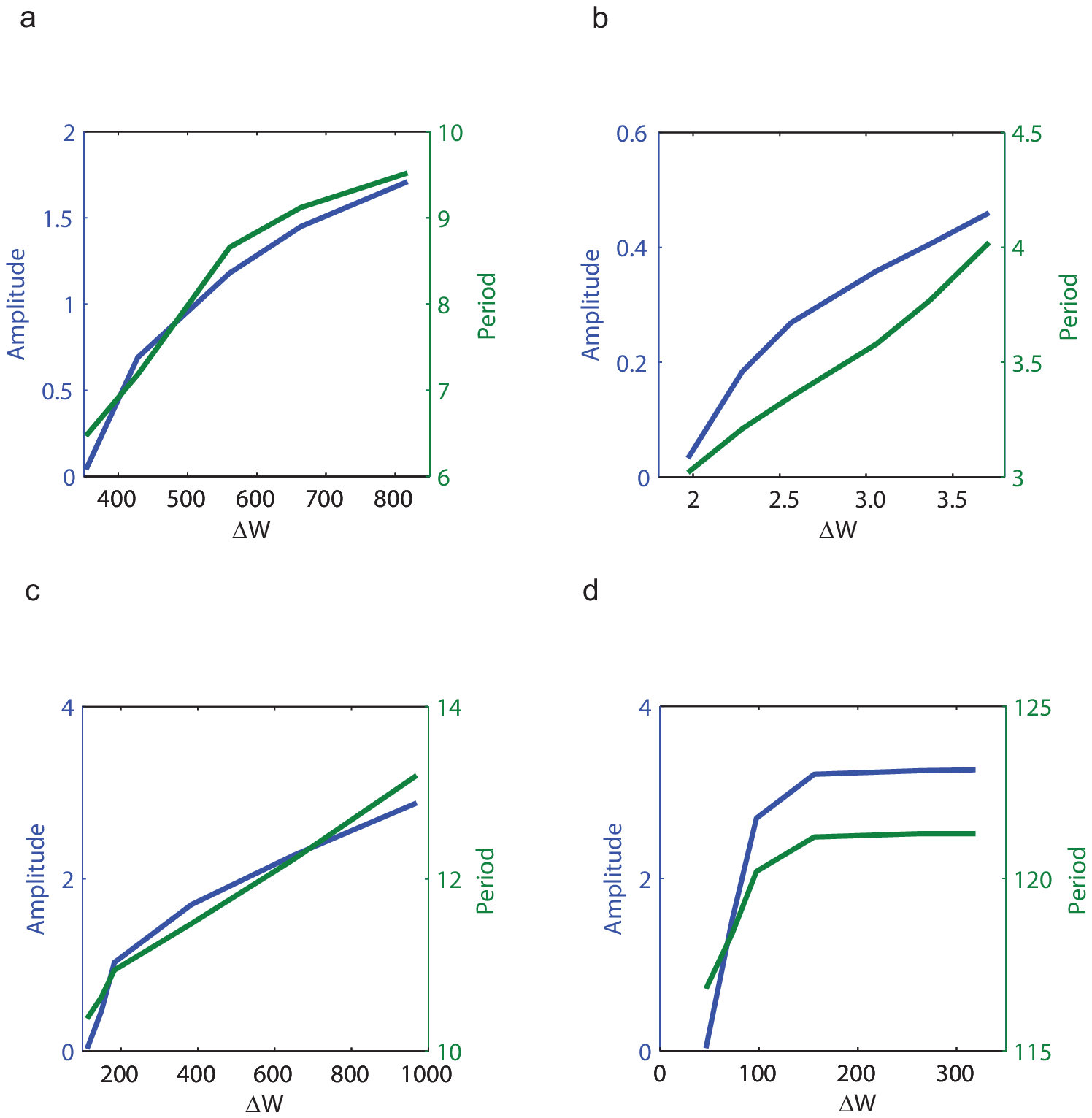}
\caption{The dependence of amplitude ($A$) and period ($T$) on energy dissipation for the four models: (a) activator-inhibitor, (b)repressilator, (c)brusselator, (d)glycolysis. For all the four cases, the critical free energy dissipation per period $W_c$ is finite at the onset of the oscillation, i.e., $A=0$. }

\end{figure}

\begin{figure}
\centering
\includegraphics{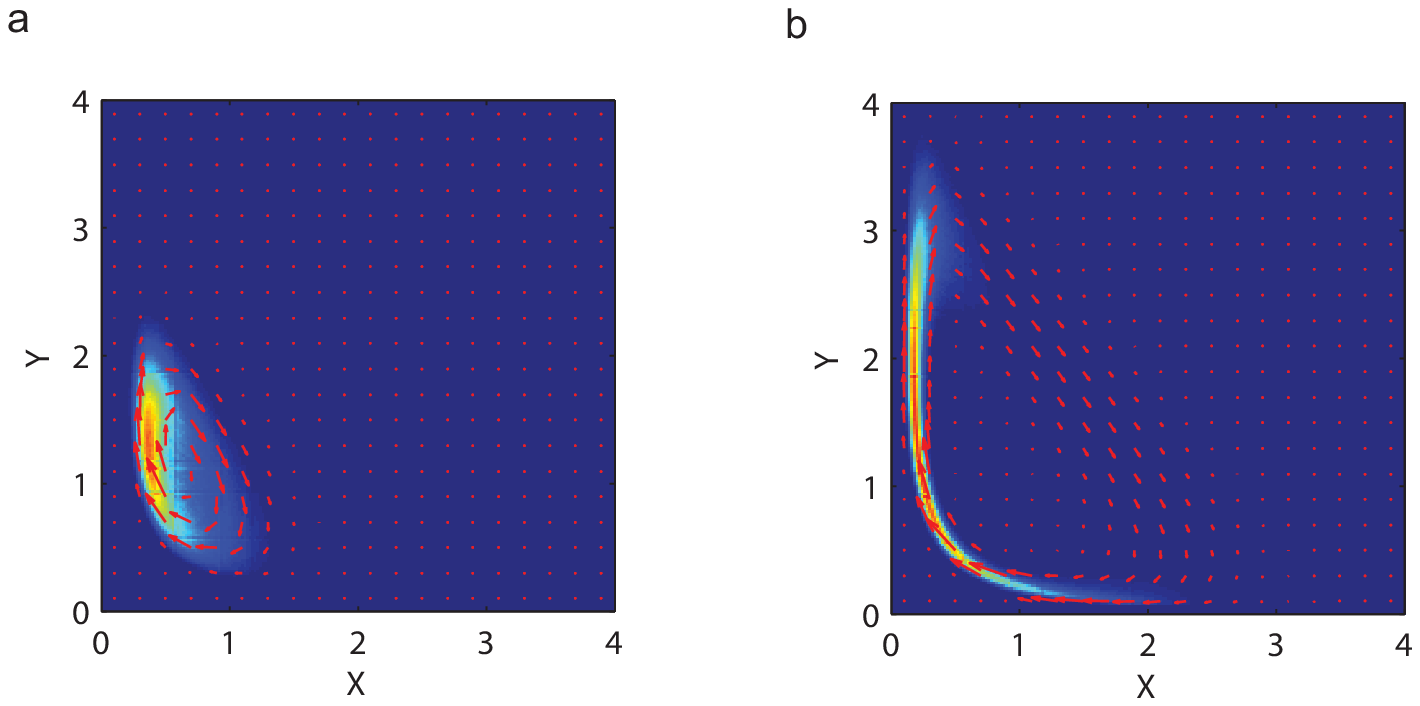}
\caption{Probability distributions and fluxes (red arrows) in the brusselator model. (a)$a=0.18$, near the onset (bifurcation point), (b) $a=0.12$, far from the bifurcation point.}

\end{figure}

\begin{figure}
\centering
\includegraphics{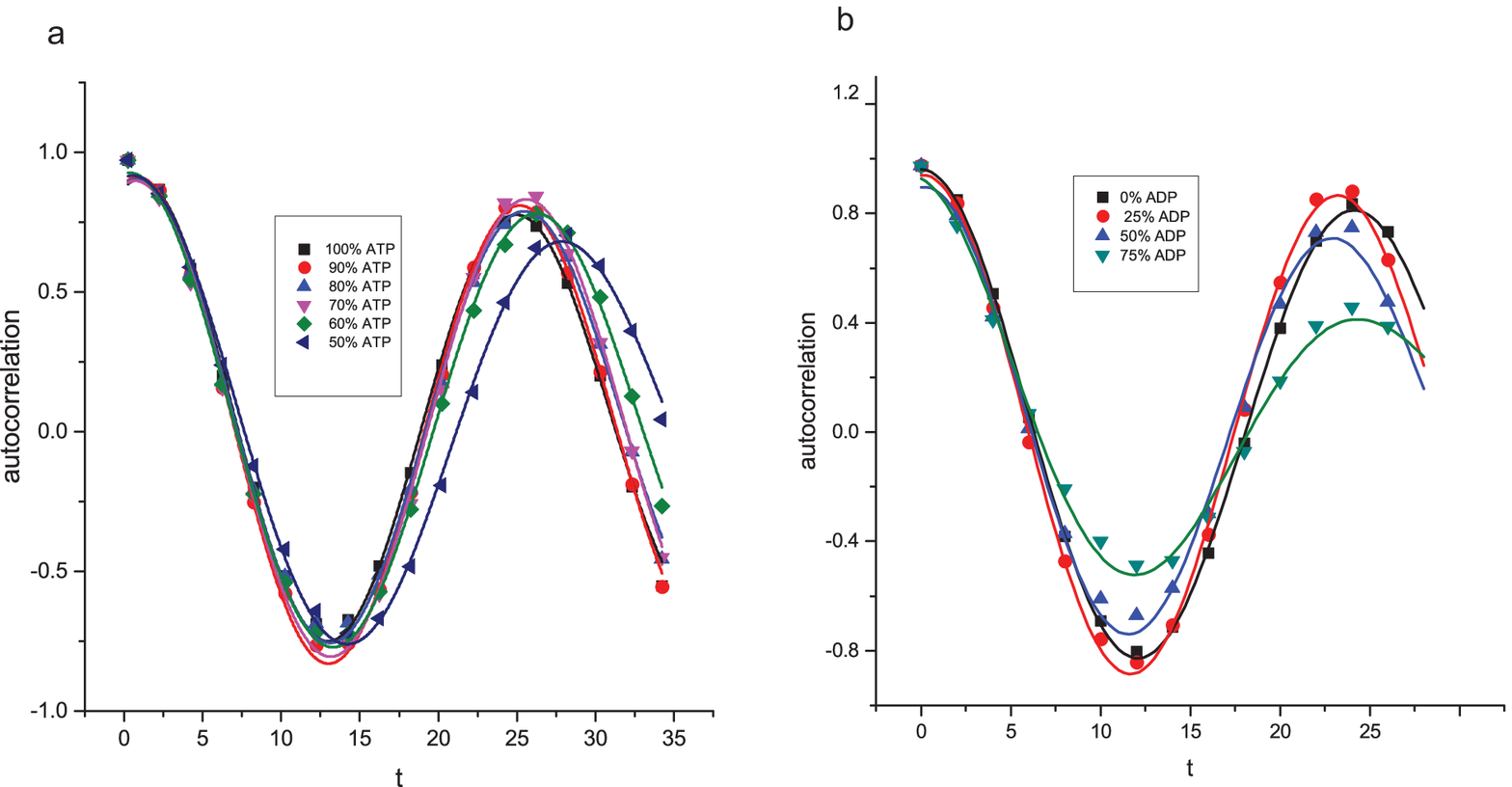}
\caption{Autocorrelation of experimental data. The autocorrelation function were calculated from the original data and fitted with $A\cos(2\pi t/T)e^{-t/\tau}$, where $T$ is the period, and $\tau$ is the correlation time. (a). Data from Ref\cite{Rust2011} with different ATP ratio. (b).Data from Ref\cite{Phong2013} with different ADP ratio.}
\label{figS_exp}
\end{figure}

\begin{figure}
\centering
\includegraphics{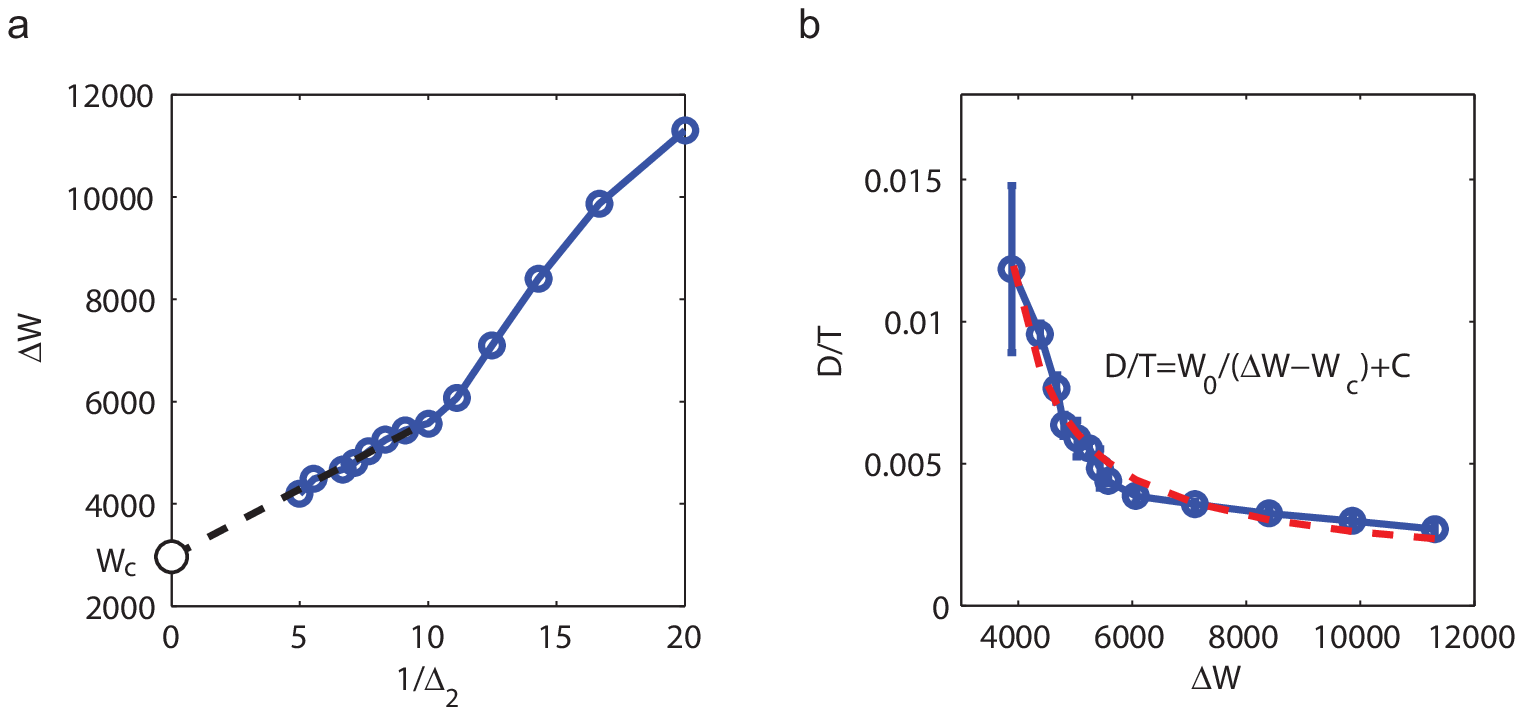}
\caption{Numerical simulation results of of the general Stuart-Landau equation (Eq. S24) with $a=1,c=1,b=2,d=1,\Delta_1=0.1$. We varied $\Delta_2\in[0.05,0.2]$. (a) Relationship between energy dissipation $\Delta W$ and  noise strength $\Delta_2$. When $\Delta_2$ is large, $\Delta W$ decreases linearly with $1/\Delta_2$. The dashed line shows when $\Delta_2\rightarrow\infty$, $\Delta W \rightarrow W_c\approx 2967$ (black circle). (b) The peak time diffusion constant $D/T$ versus $\Delta W$. The red dashed curve is the fitting inverse proportional relation, with parameters $W_0=10.2, W_c=2967,C=0.0011$.}
\end{figure}

\begin{figure}
\centering
\includegraphics{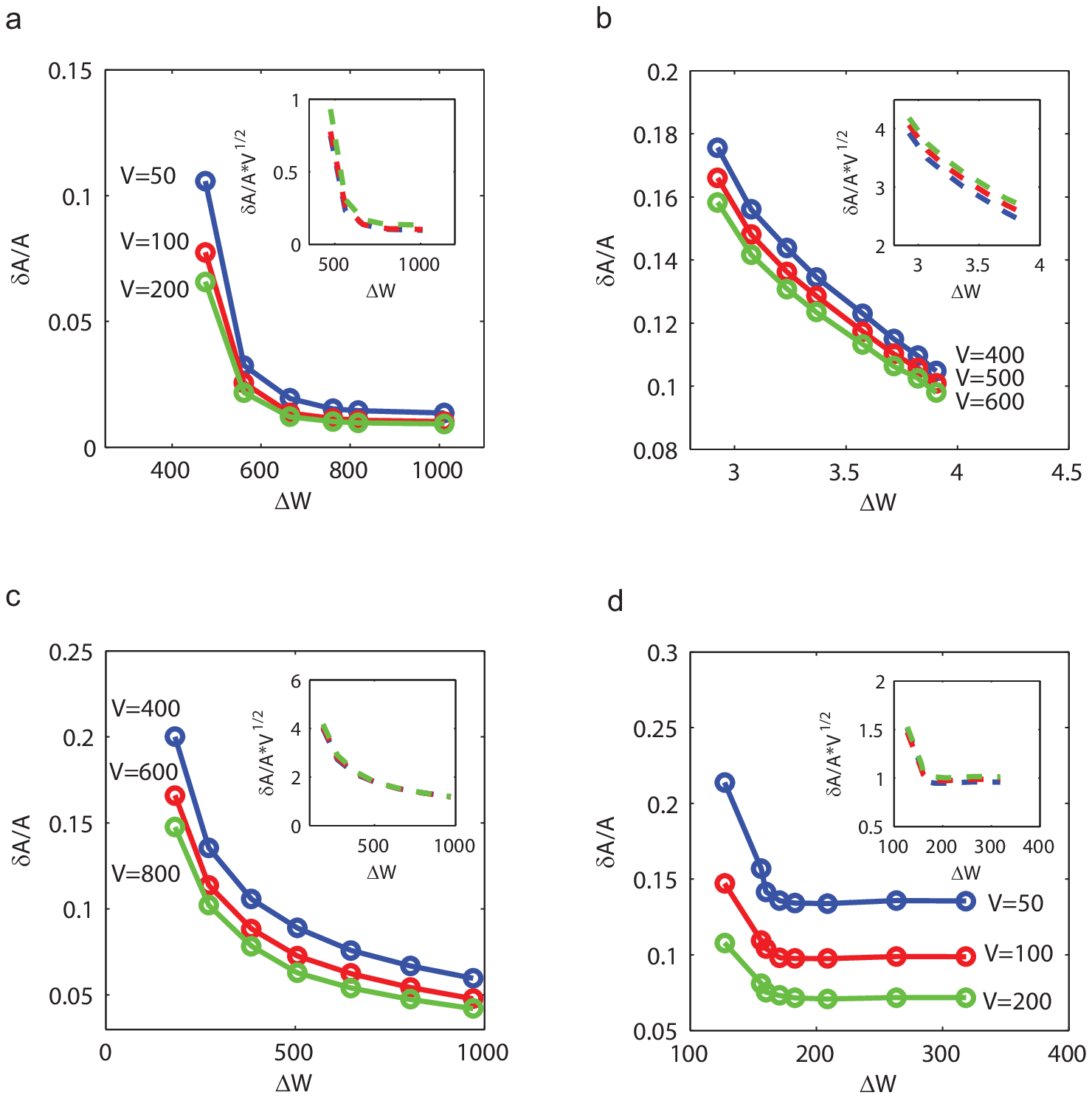}
\label{fig5}
\caption{Relation between relative amplitude fluctuation $\delta A/A$ and free energy dissipation $\Delta W$ for the four models. (a) activator-inhibitor; (b)repressilator; (c)brusselator; (d)glycolysis. Data for different volumes collapses (insets) when we scaled amplitude fluctuation with $1/\sqrt{V}$.  }
\end{figure}

\begin{figure}
\centering
\includegraphics{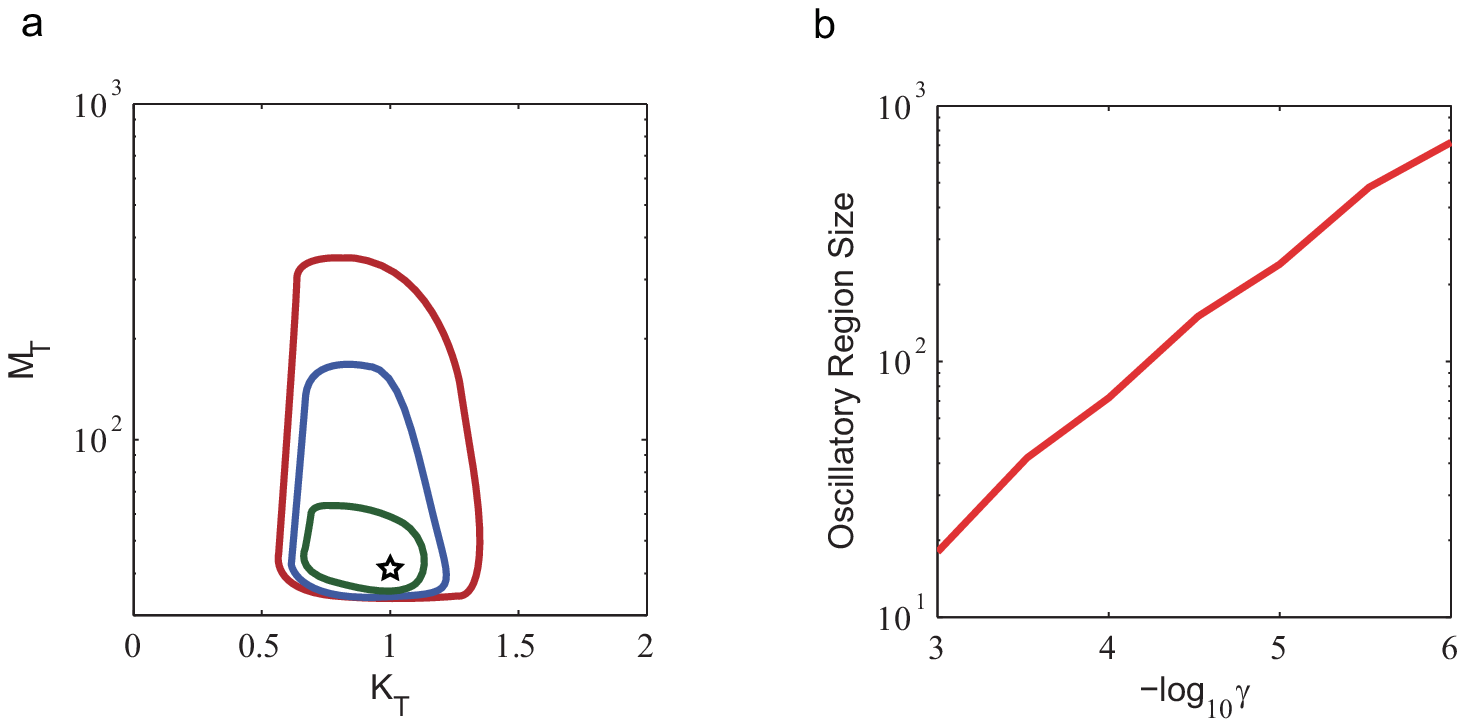}
\caption{The relationship between functional robustness and free energy dissipation. (a) The green, blue and red curves in the $(E_T, K_T)$ space correspond to the boundaries inside which oscillations exist for  $\gamma=10^{-3}$, $\gamma=10^{-4}$ and $\gamma=10^{-5}$, respectively. The star indicates the parameters in main text Fig1 a. (b) Robustness, defined as the area of oscillation in the parameter space, increases as $\gamma$ decreases. This means that higher free energy consumptions (on average) is needed for higher robustness against parameter variations in achieving oscillatory behaviors.}
\end{figure}